# Laser-Doping of Crystalline Silicon Substrates using Doped Silicon Nanoparticles

By *Martin Meseth[1], Kais Lamine[1], Martin Dehnen[1], Sven Kayser[2], Wolfgang Brock[3], Dennis Behrenberg[4], Hans Orthner[1], Anna Elsukova[5], Nils Hartmann[4], Hartmut Wiggers[1], Tim Hülser[6], Hermann Nienhaus[5], Niels Benson[1],\* and Roland Schmechel[1],\**

[1] University of Duisburg-Essen and CeNIDE, Faculty of Engineering, Bismarckstr. 81, Duisburg, Germany

[2] ION-TOF GmbH, Heisenbergstr. 15, Münster, Germany

[3] Tascon GmbH, Heisenbergstr. 15, Münster, Germany

[4] University of Duisburg-Essen and CeNIDE, Faculty of Chemistry, Universitätsstr. 5, Essen, Germany

[5] University of Duisburg-Essen and CeNIDE, Faculty of Physics, Lotharstr. 1, Duisburg, Germany

[6] IUTA, Bliersheimer Str. 58-60, Duisburg, Germany

[*]   Prof. Dr. R. Schmechel, Dr. N. Benson
E-mail: roland.schmechel@uni-due.de, niels.benson@uni-due.de
Adress: University of Duisburg-Essen, NST, Bismarckstr. 81, 47057 Duisburg, Germany
Fax: +49 (0) 203 379 3268
Phone: +49 (0) 203 379 3347, +49 (0) 203 379 1058

Abstract:

Crystalline Si substrates are doped by laser annealing of solution processed Si. For this experiment, dispersions of highly B-doped Si nanoparticles (NPs) are deposited onto intrinsic Si and laser processed using an 807.5 nm cw-laser. During laser processing the particles as well as a surface-near substrate layer are melted to subsequently crystallize in the same orientation as the substrate. The doping profile is investigated by secondary ion mass spectroscopy revealing a constant B concentration of $2 \times 10^{18}$ cm$^{-3}$ throughout the entire analyzed depth of 5 µm. Four-point probe measurements demonstrate that the effective conductivity of the doped sample is increased by almost two orders of magnitude. The absolute doping depth is estimated to be in between 8 µm and 100 µm. Further, a pn-diode is created by laser doping an n-type c-Si substrate using the Si NPs.

1. Introduction

To realize semiconductor doping a multitude of possible methods are available. The major principles are ingot doping, ion implantation, alloying using impurity metals, dopant diffusion (liquid or solid phase) using gases or doped materials deposited by epitaxy [1; 2]. However, the development of highly effective and robust low-cost doping techniques for building up pn-junctions remains challenging. While epitaxy requires vacuum, ion implantation additionally needs a temperature step to heal substrate damage and to activate the implanted dopants [3; 4]. Diffusion processes for dopant incorporation also require high temperatures. Furthermore, care has to be taken with respect to oxygen contamination [5]. The use of screen printed aluminium (Al) pastes to dope Si wafers e.g. in photovoltaics, is a low cost process as such [6; 7]. Nevertheless, for thin Si substrates used to reduce material costs, wafer bowing due to the different thermal expansion coefficients of Si and Al renders the substrates inadequate for further processing [8].

A potentially low cost doping alternative can be realized by using laser technology in combination with different doping sources such as gaseous, liquid or solid dopant materials [9 - 13]. Due to the laser radiation the temperature locally increases, leading to an incorporation of the dopants by solid-state or liquid phase diffusion [14, 15]. The type of laser used for the respective process varies with regard to wavelength and the pulse duration, ranging from the infra red to the ultra violet and from below 100 fs pulses to continuous wave radiation, respectively [12, 14 - 17].

In this article we report on a novel approach towards potentially low cost doping of crystalline Si substrates, via laser processing of highly doped Si nanoparticles (NPs). These can be produced in large quantities at a low cost via gas phase synthesis, dispersed in solvents and then deposited via spin coating or printing techniques [18].

2. Experimental
   2.1. Nanoparticle synthesis

B-doped Si NPs are synthesized using a tubular hot-wall reactor (HWR) by spray pyrolysis. The precursors monosilane and a mixture of 1 % diborane in hydrogen are fed into the reactor consisting of six heating zones and a total heating length of 1 m. A nitrogen flow serves as carrier gas. The furnace temperature is set to 1050 °C and the reactor pressure is adjusted to 400 mbar using a vacuum pump and a regulating valve. The particles are transported by the gas flow into a separator and collected on porous stainless filters. After production the material is automatically filled into plastic containers by back purging the filter. The production rate is ≈ 600 g per hour. Further information about HWR NP synthesis and the particles can be found elsewhere [18, 19]. The thickness of the oxide shell encasing the HWR NPs after air contact, is suggested to be similar to the oxide shell measured for Si NPs created in a micro wave reactor, which is in the order of 0.5 nm to 1 nm [20].

   2.2. Creation of Si-NP dispersions

By using a Netzsch MicroSeries laboratory mill, 10 wt% of Si NPs are dispersed in ethanol. In a pre-dispersion step, the NPs are milled for 1 h using 0.3 mm grinding pearls. The final dispersion step is performed using 0.1 mm grinding pearls for 2 h. After the milling, the pearls are separated from the dispersion using a 0.5 µm glass fibre filter. The particles' mean diameter is determined to be 150 nm using dynamic light scattering.

   2.3. Sample preparation

1 cm x 1 cm (100) intrinsic Si substrates (conductivity ~ 0.02 $\Omega^{-1}cm^{-1}$) of 525 µm thickness and with one polished surface are successively cleaned for 10 min in acetone and isopropanol ultra sonic baths. To remove the native $SiO_x$ on the Si pieces, the samples are then etched for 60 min in hexafluorosilicic acid at ≈ 90 °C. Subsequently the samples are washed in de-ionized $H_2O$. Si-NPs are then deposited onto the polished surface using a spin coater at 2000 rpm. The dried, as-deposited thin films exhibited a thickness of ≈ 400 nm, as measured with an Ambios XP200 profilometer. For the following laser annealing step, a near-infrared line laser with a wavelength of 807.5 nm, a spectral width of 6.8 nm and a maximum power output of 452 W is used. The laser has a top hat line profile perpendicular to its scan direction (13.38 mm FWHM) and a Gaussian profile parallel to its scan direction (52 µm FWHM). Each substrate is laser-annealed on a quartz glass piece in a chamber continuously flushed with $N_2$ or Ar during processing in order to avoid sample oxidation [21]. The glass is used to thermally insulate the substrate from the metal processing chamber and therefore to homogenize the temperature distribution in the substrate. The laser processing is conducted in two steps: The first step consists of 20 scans at maximum laser power and a scanning speed of 10 m/min in order to warm up the sample. (The warm up step was used to reduce thermal stress in the substrate. However, it is not expected to be mandatory.) For the second annealing step, the laser is scanned over the substrate at 0.3 m/min at 40 % of the maximum laser power. 200 nm thick Al contacts are then deposited via thermal evaporation at a rate of ~ 0.5 nm/min and a chamber pressure of ≈ $2 \times 10^{-6}$ mbar.

### 2.4. Structural investigation

To investigate the structure of the annealed thin film, X-ray diffraction (XRD) measurements are performed using a PANalytical X'pert Pro MPD instrument with Ni-filtered Cu K-α radiation. The data is recorded in steps of $\Delta 2\theta = 0.01$ ° over a range of 20 °≤ 2θ ≤ 120 °.

Furthermore an anisotropic etching experiment is performed. For this Si samples are immersed into diluted hydrofluoric acid in order to remove the surface oxide layer. Subsequently these are placed in a KOH solution of 70 wt% $H_2O$, 20 wt% isopropanol and 10 wt% KOH at room temperature for 5 min. For crystalline samples the surface topography changes according to the crystal orientation [22].

Characterization by transmission electron microscopy (TEM) is carried out using a 200 keV Philips, Tecnai F20 ST. For this a cross sectional lamella is cut out of the sample's surface using the focussed ion beam (FIB) of a Helios NanoLab 600 DualBeam (FEI): in a first step an area of 2 x 10 µm² is covered with a 2.5 µm thick Pt-layer to protect it from possible damage from the ion beam. The area surrounding the deposited Pt is then milled by a Ga ion beam, leaving a cross sectional lamella of 6 µm in height. The lamella is separated from the bulk and thinned down to a final width of ≈ 80 nm. During the latter process the ion beam current is decreased, in order to reduce the amount of residual Ga on the faces of the lamella.

### 2.5. Elemental analysis

To analyse the elemental composition of the samples, time-of-flight secondary ion mass spectroscopy (TOF SIMS) measurements are performed using a TOF-SIMS[5] instrument by ionTOF in the "Dual Beam Depth Profiling" mode, with sputter beams of $O_2^+$ or $Cs^-$. The beams are used at a low energy in order to avoid forward implantation of B, and to ensure a high sensitivity for oxygen or oxygen containing molecules and B during the analysis. The respective sputter conditions are: 2 keV ion energy, 343 nA target current for $O_2^+$ primary ions and 0.5 keV ion energy, 32 nA target current for $Cs^-$ primary ions. A sputter area of 300 x 300 µm² is used for both primary ions. For the analysis beam $Bi_3^+$ primary ions are used at ion energies of 30 keV and target currents of 0.14 pA or 30 pA for the respective negative and positive polarity of the analyzer on an area of 100 x 100 µm² within the sputter area, in order to avoid edge contamination. $Bi_3^+$ is chosen for the analysis beam in order to maximize the mass resolution.

## 2.6. Electrical characterization

For the electrical 4-point characterization, Al contacts are deposited on top of the substrates, cf. inset of **fig. 5**. Using a Keithley 4200 SCS, the conductance is obtained from the voltage difference U between the inner electrodes and the constant applied current I, which is driven through the outer electrodes. Considering the spacing and overlapping width of the electrodes S = 0.1 mm and W = 7.4 mm, respectively, a conductivity σ can be determined by:

$$\sigma = \frac{I}{U} \cdot \frac{S}{W \cdot d} \qquad (1)$$

where d is the thickness of the sample.
Electro-optical characterization is carried out using a Keithley 238 SMU in combination with an AM1.5G Wacom SolarSimulator.

## 3. Results and Discussion

The insets in **figure 1** a) and b) show photographs of as deposited samples and after subsequent laser annealing, respectively. After annealing the reflectivity of the coated surface is strongly increased and its colour is changed from brown to silver, indicating a smooth surface. This is confirmed by cross sectional SEM pictures of the samples prior to and after laser annealing, cf. fig. 1: While the as-deposited thin film consists of flake-shaped particles and is highly porous, after laser treatment it is hardly possible to distinguish between the laser annealed, surface-near layer and the bulk of the Si substrate. From the missing pores strong compaction is evident, which is assumed to be the result of the melting of particles and a partial melting of the substrate. Therefore, the border between the substrate and the thin film of NPs has vanished.

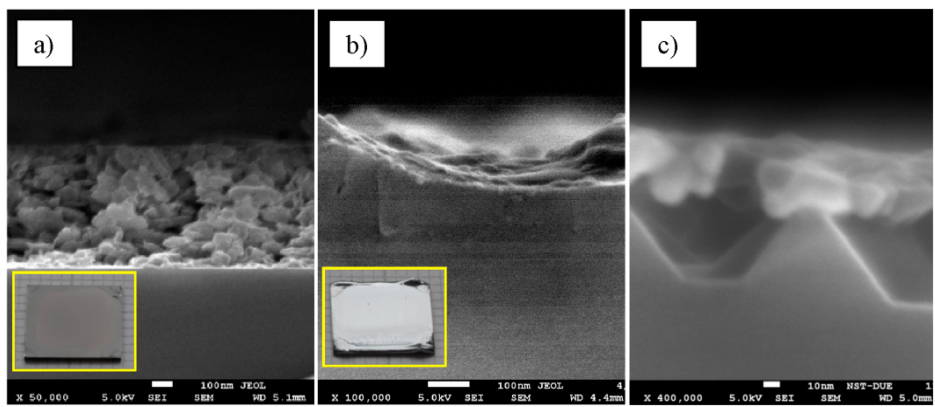

**Figure 1**
a) Cross sectional SEM image of a wafer piece after NP deposition. A clear border between the wafer and the coating is visible. b) After the laser treatment the pores between particles as well as the wafer / thin film boundary have vanished. The insets in a) and b) show the respective top view images of an as-deposited and a laser annealed sample. c) The cross section after wet etching of the laser annealed sample in aqueous KOH shows a pyramidal structure in agreement with a crystalline (100)-surface orientation.

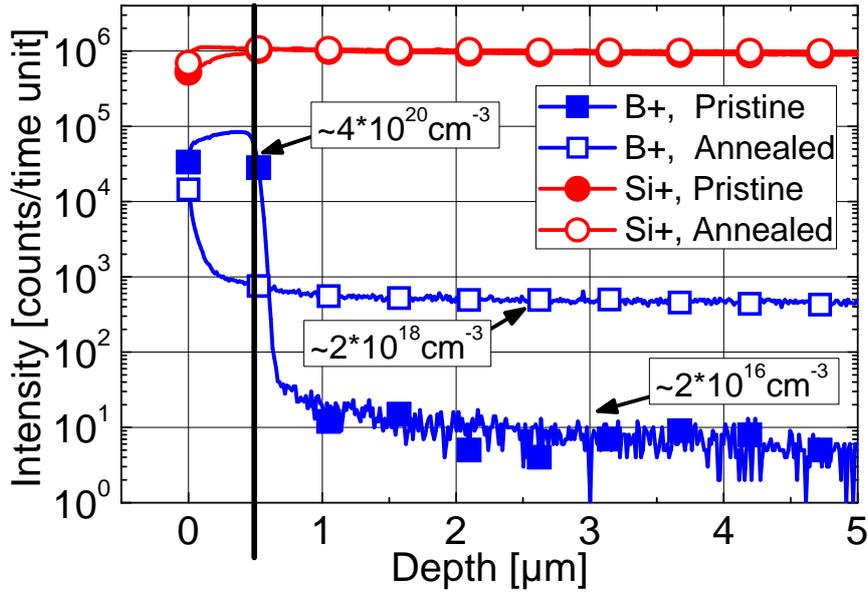

**Figure 2**
TOF-SIMS data: secondary ion counts per unit time plotted against the sputtering depth for boron (squares) and silicon (circles) before (filled symbols) and after (open symbols) laser treatment. The values given in the diagram refer to the effective boron concentration of the samples.

3.1. SIMS measurements

To investigate whether B diffused into the Si substrates during the laser processing step, annealed samples are analyzed using TOF-SIMS. **Fig. 2** depicts the depth profiles of B and Si concentration prior to and after laser processing. While the Si distribution is almost unaffected by the annealing step, a significant change is obtained for the B distribution: prior to laser annealing the B concentration is more or less constant at $7\times10^4$ ($\pm 1\times10^4$) counts per unit time for a sputter depth of up to 500 nm. For sputter depths exceeding 500 nm the B concentration rapidly drops below 10 counts per unit time. Using a relative sensitivity factor determined for B in a Si matrix, the determined values can be ascribed to B concentrations of $\approx 4\times10^{20}$ cm$^{-3}$ for d < 500 nm and $\approx 2\times10^{16}$ cm$^{-3}$ for d > 500 nm. These results are in good agreement with the initial doping concentration of the Si NPs ($\sim 10^{21}$ cm$^{-3}$) and the detection limit of the TOF-SIMS ($\sim 1\times10^{16}$ cm$^{-3}$), respectively. Further, the observed transition from the doped NPs film to the undoped Si substrate at a depth of $\approx 500$ nm is in line with the determined as deposited NP thin film thickness. The difference in thickness of $\approx 100$ nm is ascribed to a TOF-SIMS measurement inaccuracy, because of the porous nature of the thin film.

After the laser annealing, the B concentration drops within the topmost 200 nm as a function of the reciprocal depth down to a value corresponding to $\approx 4\times10^{18}$ cm$^{-3}$ and then levels at $\approx 2\times10^{18}$ cm$^{-3}$. This value only marginally decreases further over the sampling depth of 5 μm, substantiating the successful doping of the Si substrate within a depth of at least 5 μm. A three dimensional plot of the SIMS measurement (not shown here) does not exhibit B agglomerates, suggesting a homogeneous doping distribution realized by the laser annealing. However, we cannot exclude, dopants to be electrically inactive.

Considering the low diffusivity of B in crystalline silicon (D = $1\times10^{-12}$ cm$^2$ s$^{-1}$ at 1200 °C and at a background boron concentration of $1\times10^{19}$ cm$^{-3}$) and the comparatively short laser processing time (t < 1 min), it is concluded that the observed B doping profile cannot be the result of diffusion in crystalline Si [23]. Therefore, the laser annealing step is concluded to result in a melted substrate surface layer leading to B diffusion in liquid Si with a diffusion constant of D = 2.4x10$^{-4}$ cm$^2$ s$^{-1}$ [24]. In addition, convection inside the melt may contribute to the doping process, resulting in the discussed homogeneous doping profile. In particular, the resulting temperature difference between the liquid and the solid silicon phase may act as the driving force for a convection mechanism [25].

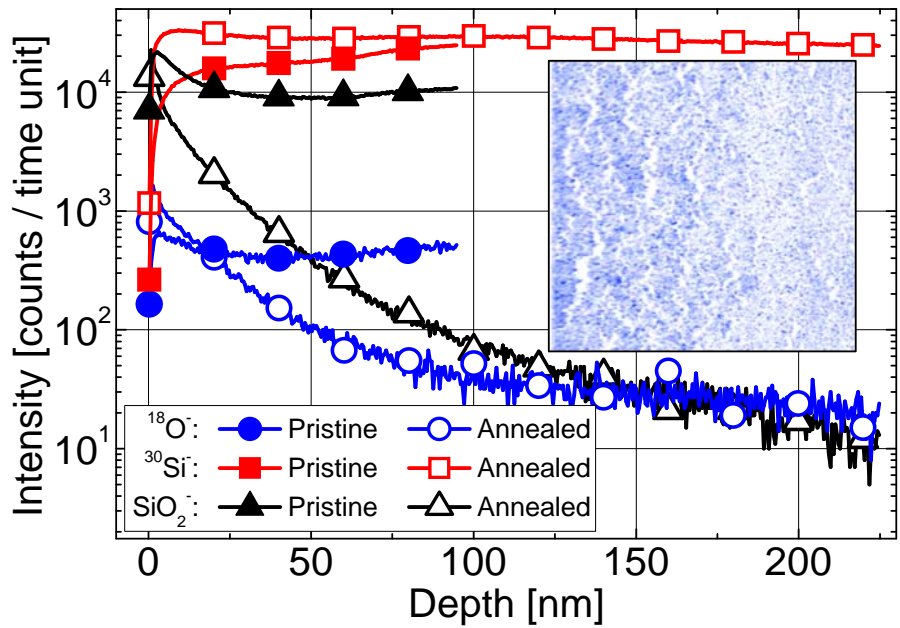

**Figure 3**
TOF-SIMS-data showing the secondary ion counts per unit time plotted against the sputtering depth. The amount of oxygen (triangles) and SiO$_2$ (squares) is reduced by the laser treatment. The inset shows a surface scan for oxygen containing species (blue colour) and Si (white).

In order to investigate the distribution of oxygen containing species in the sample structure, a sample is analyzed using Cs$^-$ primary ions. The result is shown in **fig. 3**, comparatively illustrating the depth profiles of Si, O and SiO$_2$ as representatives for oxidized Si species. Here only relative concentrations are discussed; an oxygen standard for calibration has not been available during the time of the analysis. As discussed regarding fig. 2, the Si concentration does not change during laser processing. In contrast, the profile of SiO$_2$ and oxygen change from a more or less homogeneously distribution in the as-deposited NP thin film to an exponential decrease by almost two orders of magnitude within the analyzed sputtering depth of 225 nm. Considering the relative signal intensities of SiO$_2$ and oxygen normalized to Si before and after laser annealing, it is found, that oxygen containing species are reduced by a factor of ≈ 5-6 due to the laser annealing, which is suggested to be the result of the formation of volatile compounds during laser processing such as N$_x$O$_y$, SiO$_x$(g) or O$_x$. Taking into account the rather homogeneous elemental distribution in the as-deposited thin film, in comparison to the strong decay of the SiO$_2$ and the oxygen signals in the laser-processed samples, we propose that the oxygen containing species accumulate at the topmost surface of the sample during laser processing. This however does not lead to a homogeneous surface layer of SiO$_x$, as evident from the inset in fig. 3. Here an elemental map of the analyzed surface is displayed, showing oxygen rich (coloured) and Si rich areas (white).

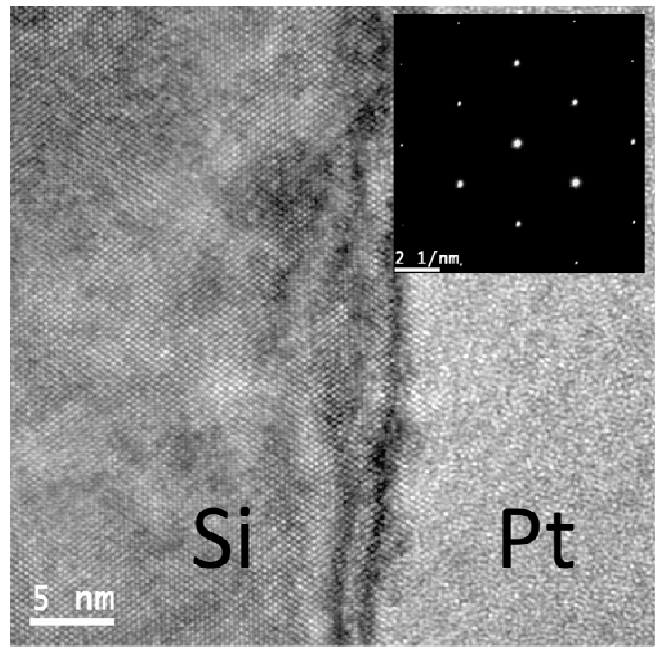

**Figure 4**
High-resolution transmission electron microscope (TEM) image of a surface-near layer: the amorphous Pt-layer on the right is due to the FIB preparation. The Si-sample is on the left, displaying a periodic structure of three fold symmetry. A transmission electron diffraction (TED) pattern of this thin film is presented in the inset showing the spots of (110)-oriented Si.

### 3.2. Structural investigations

To investigate the crystallinity of the laser annealed thin film, XRD measurements (not shown here) are performed. The diffraction pattern shows a {400}-peak at 2θ = 69.7 °; no other peaks are visible. However, thin or amorphous films are hard to analyze using XRD measurements, since the signal can originate from the substrate only. Therefore, a complementary etching experiment in aqueous KOH solution is performed. Wet etching of crystalline Si in KOH is anisotropic. Hence, respective etching profiles provide information on the crystallinity. B doping at high concentrations inhibits the etching process. For this reason, an intrinsic (100) Si wafer, is coated with intrinsic Si NPs, laser annealed using the same parameters as discussed above and etched in KOH-solution. Subsequently the sample's cross section is analysed via SEM and the resulting micrograph is illustrated in fig. 1 c). Evidently etching yields pyramidal structures which are typical for (100) oriented Si samples [22]. It is therefore reasonable to conclude that the laser annealed thin film has re-crystallized in the same orientation as the substrate. The granular layer on top of the pyramidal structures is attributed to residual silicon oxide, which remained on the surface after etching in hydrofluoric acid.

To prove the structure of the re-solidified thin film, a cross sectional lamella of the surface parallel to the wafer flat, i.e. parallel to the (110)-direction, is prepared by focussed ion beam (FIB) and investigated using high-resolution transmission electron microscopy (TEM). In **fig. 4** a TEM micrograph shows, from right to left, an amorphous Pt-layer, which originates from FIB-preparation, and the substrate. The whole re-solidified, surface-near layer of the substrate exhibits a periodic, three-fold symmetric structure, as it is found for c-Si. To clarify the crystallinity of this layer, transmission electron diffraction (TED) measurements are performed. The diffraction pattern is shown in the inset of fig. 4, displaying the characteristic spots for (110)-oriented Si [26]. By changing the position of the analyzed area on the lamella, the same pattern is found for the whole re-solidified layer. This confirms that the melt crystallizes in the same orientation as the substrate.

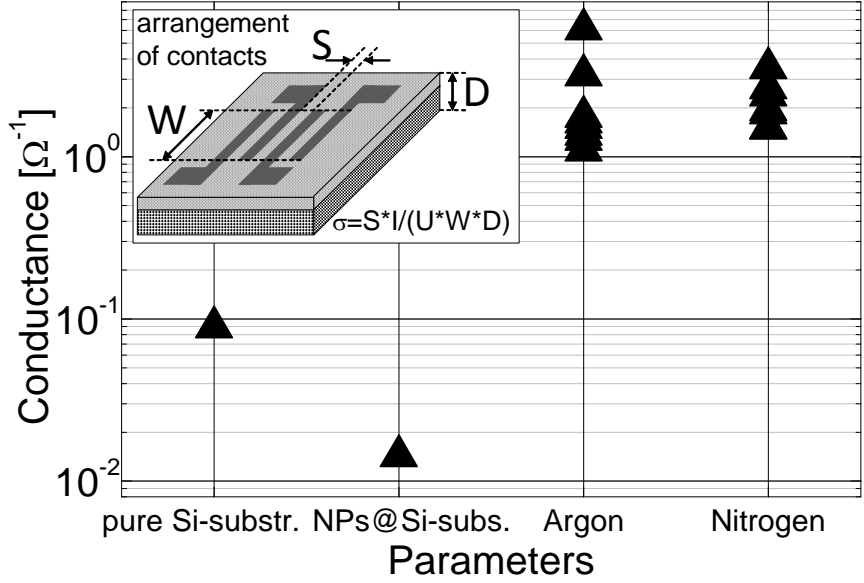

**Figure 5**
Conductivity measurement results for i-Si substrates, i-Si + NP and i-Si + NP + laser annealing in argon and nitrogen. The inset shows the geometry of electrode contacts on the sample: grey = aluminium contacts; facetted area = laser doped surface region; chequered area = unchanged wafer substrate.

### 3.3. Conductance measurements

To characterize the electrical effect of doping, the conductances of the samples are determined by 4-point probe measurements. Illustrated in **fig. 5** are the measured results as black triangles. For clean Si-substrates the conductance is obtained to be 0.089 $\Omega^{-1}$. After NP deposition, sample conductances in the order of 0.014 $\Omega^{-1}$ are measured. This decrease can be explained by the granular structure and by the Si-SiO$_x$ core-shell nature of the as-deposited NP film, which impedes the electrical current [27]. After laser crystallization of the NP thin film, the conductance can be improved up to a maximum value of 3.5 $\Omega^{-1}$. Using eq. (1) in combination with the entire sample thickness of $d_{wafer}$ = 525 µm, an effective sample conductivity of $\sigma_{eff}$ = 1.5 $\Omega^{-1}$ cm$^{-1}$ is obtained. According to Bulucea et al. [28] this conductivity corresponds to an effective doping concentration of ≈ 2x10$^{16}$ cm$^{-3}$. No significant change of the thin film conductivity is observed by varying the flush gas N$_2$ vs. Ar.

### 3.4. Estimation of the doping depth

An upper limit for the doping depth can be obtained from the SIMS measurement, assuming that the total amount of boron remains the same before and after laser annealing. If the SIMS-doping profile is approximated by a box function, the relation:

$$c_{init} \cdot d_{init} = const = c_{final} \cdot d_{final} \qquad (2)$$

leads to $d_{final}$ = 100 µm. Here $c_{init}$ and $d_{init}$ represent the doping concentration and the thickness of the as-deposited Si-NP thin film, whereas $c_{final}$ and $d_{final}$ represent the homogeneous doping concentration and dopant diffusion depth after the laser annealing step, respectively. However, since the SIMS signal may be inaccurate by a factor of two or higher, the estimate for $d_{final}$ is expected to have a similar uncertainty, cf. eq. (2).

A lower limit for the diffusion depth is obtained from the conductivity. For this purpose a two layer structure is assumed (see fig. 5). One layer represents the B-doped thin film of unknown thickness $d_B$ and conductivity $\sigma_B$. The second layer represents the unmodified, intrinsic substrate with a thickness of $d_{intr} = d_{wafer} - d_B = 525$ nm $- d_B$ and a conductivity of $\sigma_{intr} = 0.022$ $\Omega^{-1}$ cm$^{-1}$ (see above). The conductivity measured effectively ($\sigma_{eff}$) contains both contributions:

$$d_{wafer} \cdot \sigma_{eff} = \sigma_B \cdot d_B + \sigma_{intr} \cdot d_{intr}$$

which yields for the doping depth:

$$d_B = d_{wafer} \frac{\sigma_{eff} - \sigma_{intr}}{\sigma_B - \sigma_{intr}}$$

The conductivity of the doped region $\sigma_B$ is estimated to be $\sigma_B \approx 33$ $\Omega^{-1}$ cm$^{-1}$ by comparing the doping concentration of $2 \times 10^{18}$ cm$^{-3}$ obtained from the SIMS measurement with reference 28. This results in a mean value of $d_B \approx 8$ µm.

Since both estimations are based on several assumptions leading only to minimum and maximum values, the real doping depth is therefore expected to be in between $d_B$ and $d_{final}$, thus between 8 µm and 100 µm.

### 3.5. Creation and characterization of a pn-junction

To investigate if laser processing of Si NPs on a Si substrate allows to change the type of majority doping, a pn-junction is realized by the following experiment. B-doped Si NPs are spin coated onto a one side-polished, (100), 1.5 cm x 1.5 cm c-Si substrate exhibiting a phosphorus doping concentration of $3 \times 10^{17}$ cm$^{-3}$ and a thickness of 525 µm. After laser crystallization 200 nm thick Al contacts of 9 mm x 9 mm are formed in the centre of both surfaces by physical vapour deposition. The current voltage (I(U)) characteristic of the sample as measured in the dark is illustrated in **fig. 6** (black triangles), exhibiting a rectifying behaviour with an on / off ratio of 207 (at U = +/- 1 V) and a reverse current of I ≈ 1 mA at U = - 1 V. Using a standard equivalent one diode circuit as depicted in the inset a) of fig. 6, the I(U) characteristic can be described using equation (3):

$$I(U) = \frac{U_{shu}}{R_{shu}} + I_{diode} \cdot \left( \exp\left[ e \frac{U_{shu}}{n \cdot k_B \cdot T} \right] - 1 \right) = \frac{U - I \cdot R_{ser}}{R_{shu}} + I_{diode} \cdot \left( \exp\left[ e \frac{U - I \cdot R_{ser}}{n \cdot k_B \cdot T} \right] - 1 \right)$$
(3)

Here U and $U_{shu}$ are the applied voltage and the voltage drop over the shunt resistance $R_{shu}$, T represents the absolute temperature, $R_{ser}$ is the serial resistance, $I_{diode}$ and n are the saturation current and the ideality factor of the diode and e and $k_B$ are the elementary charge and the Boltzmann constant, respectively. The resulting fit is in good agreement with the experimental data, cf. dashed line in fig. 6.
Using this fit a series resistance of $R_{ser}$ = 1.3 Ω and a shunt resistance of $R_{shu}$ = 1100 Ω can be extracted. To be application relevant, according to Goetzberger et al. [5], the value of $R_{ser}$ needs to be reduced below ≈ 0.75 Ω , whereas $R_{shu}$ is already in a good range.
According to the equivalent circuit the reverse current is the sum of the diode's reverse leakage current $I_{diode}$ = 20.0 µA and the current through the shunt resistance. The obtained high reverse current value of I ≈ 1 mA at U = -1 V is therefore predominated by the low shunt resistance. Further, an ideality factor of n = 3 can be extracted from the fit. This value exceeds the Sah-Noyce-Shockley limit (1 ≤ n ≤ 2) [29]. However, according to Shah et al. [30] ideality factors with n > 2 are possible if additional non-linear elements, like a Schottky electrode / semiconductor contact or additional defect based transport paths, are present in the circuit. Especially the latter seems likely here due to the high oxygen content.
The diode is further characterized using standard AM1.5 irradiation. The resulting current voltage characteristic is illustrated in fig. 6 (green squares). A photovoltaic effect is obtained, with an open circuit voltage of $U_{oc}$ = 0.438 V, a short circuit current of $I_{sc}$ ~ 8 mA and a fill factor of FF = 51.2 %. Using a conservative approximation, the photoactive area of the device is calculated by subtracting the contact area from the substrate area, resulting in a short circuit current density of $J_{sc}$ = 5.334 mA/cm² and a low efficiency of η ≈ 1.2 %. The low photovoltaic performance is the consequence of this proof of principle experiment and its non-ideal device design. To make this concept applicable to a photovoltaic application, the emitter thickness needs to be reduced, for example by the use of different laser systems (shorter wavelength, pulsed operation). Furthermore a more adequate doping level needs to be realized, for example by mixing intrinsic and doped NP dispersions (digital doping, [31]), by reducing the doping concentration of the primary NPs or by using thinner NP thin films. Further, care has been taken for an oxide free processing.

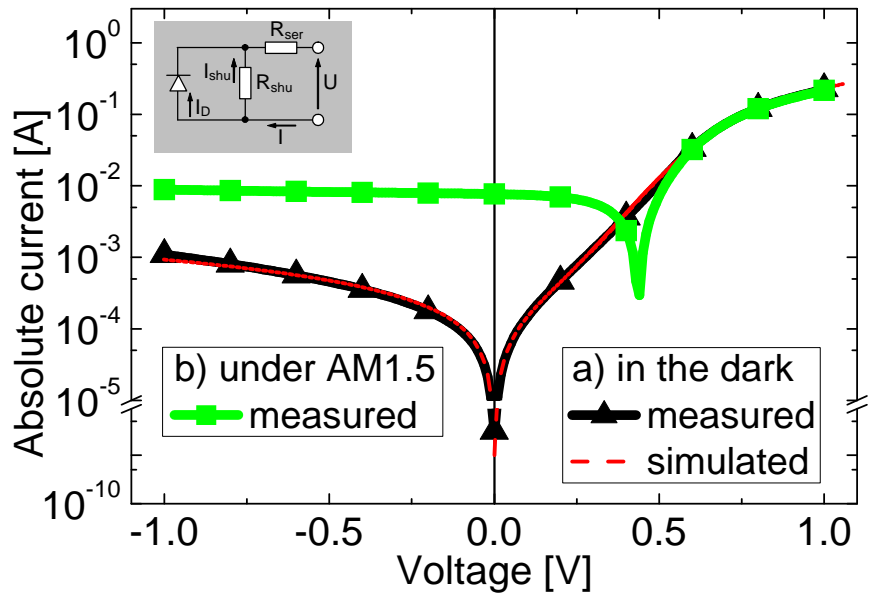

**Figure 6**
I(U) characteristic of a pn-sample realized by laser doping of an n-type Si substrate with p-type NPs: black triangles represent the dark characteristic, green squares the characteristic under AM1.5 irradiation. The red dashed and blue dotted lines represent simulated data based on equivalent circuits shown by the insets a) and b).

4. Summary

In conclusion, the laser crystallization of p-type Si nanoparticles (NPs) on intrinsic c-Si-substrates using a near-infrared line-laser is demonstrated. Though no particular oxide removal step has been applied prior to laser processing, the substrate's conductivity is increased by more than two orders of magnitude up to 1.5 $\Omega^{-1}$ cm$^{-1}$. As confirmed by SIMS measurements, substrate doping of up to $\approx 2\times10^{18}$ cm$^{-3}$ is feasible employing this process. We anticipate that during the laser processing the Si NPs, as well as a surface-near layer of the substrate are melted. Hence, deep penetration of the dopant into the substrate is attributed to diffusion, potentially enhanced by a convection process. Additionally, it is demonstrated that laser annealing reduces the oxygen content in the modified thin film by a factor of $\approx$ 5-6. XRD and TEM measurements as well as an anisotropic etching experiment lead to the conclusion, that the re-solidified film is crystalline and exhibits the same orientation as the substrate. Using a proof of principle pn- diode structure, it is demonstrated that the substrate doping type can be changed. The diode exhibits a clear rectifying behaviour with an on / off ratio of 207. However, in order to be able to evaluate the potential of this technology for photovoltaic or electronic device applications in general, further application specific process optimization needs to be conducted.

*Acknowledgements*
We gratefully acknowledge the financial support by the DFG (Research Training Group 1240 – Nanotronics), and by the Ministry of Innovation, Science and Research of the State of North Rhine-Westphalia, Germany within in the NanoEnergieTechnikZentrum (NETZ, Objective 2 Programme: European Regional Development Fund, ERDF) framework and a research scholarship for rollable solar cells. For laser support we thank the Lissotschenko Mikrooptik GmbH (LIMO). For assistance with DLS and XRD measurements we thank the research group NPPT of M. Winterer (Duisburg) with special thanks to A. Sandmann and A. Kompch, respectively. TOF-SIMS measurements at the Tascon GmbH and ionTOF GmbH are gratefully acknowledged. We furthermore thank M. Bartsch and A. Beckel from the experimental physics research group of A. Lorke (Duisburg) for FIB preparation. For fruitful TED discussions the authors thank Max Klingsporn from IHP GmbH. For financial support of TEM measurements we thank ICAN.